\documentclass[12pt]{article}
\usepackage{graphics} 
\usepackage{cite}
\usepackage{epsfig}
\newcommand{\beq}{\begin{equation}}
\newcommand{\eeq}{\end{equation}}
\newcommand{\beqa}{\begin{eqnarray}}
\newcommand{\eeqa}{\end{eqnarray}}
\newcommand{\bea}{\begin{eqnarray}}
\newcommand{\eea}{\end{eqnarray}}

\newcommand   {\ev}[1]   {\langle #1\rangle}
\newcommand   {\Ca}      {C${}_{\alpha}$}
\newcommand   {\Cb}      {C${}_{\beta}$}
\newcommand   {\Cd}      {C${}_{\delta}$}
\newcommand   {\Cp}      {C${}^{\prime}$}

\newcommand   {\Eloc}    {E_{\mbox{{\scriptsize loc}}}}
\newcommand   {\Esa}     {E_{\mbox{{\scriptsize sa}}}}
\newcommand   {\Ehb}     {E_{\mbox{{\scriptsize hb}}}}
\newcommand   {\Eihb}    {e_{\mbox{{\scriptsize hb}}}}
\newcommand   {\Ecol}    {E_{\mbox{{\scriptsize col}}}}
\newcommand   {\Tc}      {T_{\mbox{{\scriptsize c}}}}

\newcommand   {\ehb}     {\epsilon_{\mbox{{\scriptsize hb}}}}
\newcommand   {\ecol}    {\epsilon_{\mbox{{\scriptsize col}}}}

\newcommand   {\shb}     {\sigma_{\mbox{{\scriptsize hb}}}}
\newcommand   {\scol}    {\sigma_{\mbox{{\scriptsize col}}}}

\newcommand   {\Rg}      {R_{\mbox{{\scriptsize g}}}}

\newcommand  {\ACR}      {{\it Acc. Chem. Res.\ }}

\newcommand  {\Bioch}    {{\it Biochemistry\ }}

\newcommand  {\EL}       {{\it Europhys.\ Lett.\ }}

\newcommand  {\JCP}      {{\it J.\ Chem.\ Phys.\ }}
\newcommand  {\JMB}      {{\it J.\ Mol.\ Biol.\ }}

\newcommand  {\Nat}      {{\it Nature\ }}
\newcommand  {\NSB}      {{\it Nat.\ Struct.\ Biol.\ }}

\newcommand  {\Pro}      {{\it Proteins\ Struct.\ Funct.\ Genet.\ }}
\newcommand  {\ProEng}   {{\it Protein\ Eng.\ }}
\newcommand  {\ProSci}   {{\it Protein\ Sci.\ }}

\newcommand  {\PNAS}     {{\it Proc.\ Natl.\ Acad.\ Sci.\ USA\ }}

\newcommand  {\Sci}      {{\it Science\ }}

\newcommand  {\TBS}      {{\it Trends Biochem. Sci.\ }}

\begin{document}

\begin{flushright}
LU TP 01-24\\
November 5, 2001
\end{flushright}

\vspace{0.4in}

\begin{center}

{\LARGE \bf Folding of a Small Helical Protein} 

{\LARGE \bf Using Hydrogen Bonds and} 

{\LARGE \bf Hydrophobicity Forces}

\vspace{.6in}

\large
Giorgio Favrin, Anders Irb\"ack and Stefan 
Wallin\footnote{E-mail: favrin,\,anders,\,stefan@thep.lu.se}\\   
\vspace{0.10in}
Complex Systems Division, Department of Theoretical Physics\\ 
Lund University,  S\"olvegatan 14A,  S-223 62 Lund, Sweden \\
{\tt http://www.thep.lu.se/complex/}\\

\vspace{0.3in}	

Submitted to \Pro

\end{center}
\vspace{0.2in}
\normalsize
Abstract:\\
A reduced protein model with five to six atoms per amino acid and 
five amino acid types is developed and tested on a three-helix-bundle
protein, a 46-amino acid fragment from staphylococcal protein A. 
The model does not rely on the widely used G\=o approximation where 
non-native interactions are ignored. We find that the collapse transition 
is considerably more abrupt for the protein A sequence than for 
random sequences with the same composition. The chain collapse is found
to be at least as fast as helix formation. Energy minimization 
restricted to the thermodynamically favored topology gives a structure 
that has a root-mean-square deviation of 1.8~\AA\ from the native structure. 
The sequence-dependent part of our potential is pairwise additive. Our
calculations suggest that fine-tuning this potential by parameter
optimization is of limited use.   

\newpage

\section{Introduction}

In recent years, several important insights have been gained into 
the physical principles of protein folding~\cite{Sali:94,Bryngelson:95,
Dill:97,Klimov:98,Nymeyer:98,Hao:98}.
Still, in terms of quantitative predictions, it is clear that it
would be extremely useful to be able to perform more realistic 
folding simulations than what is currently possible. In fact,
most models that have been used so far for statistical-mechanical
simulations of folding rely on one or both of two quite drastic
approximations, the lattice and G\=o~\cite{Go:78} approximations.        

The reason that lattice models have been used to 
study basics of protein folding is partly computational,
but also physical --- on the lattice, it is known what
potential to use in order for stable and fast-folding 
sequences to exist (a simple contact potential is 
sufficient). How to satisfy these criteria for 
off-lattice chains is, by contrast, largely unknown, 
and therefore many current off-lattice
models~\cite{Zhou:97,Nymeyer:98,Shea:98,Zhou:99,Shea:99,
Clementi:00a,Clementi:00b,Shimada:01} use G\=o-type 
potentials~\cite{Go:78} where non-native interactions 
are ignored. The use of the G\=o approximation has some 
support from the finding that the native structure is a 
determinant for folding kinetics~\cite{Plaxco:98,Baker:00}. 
However, it is an uncontrolled approximation, and it is, 
of course, useless when it comes to structure prediction, 
as it requires prior knowledge of the native structure. 

In this paper, we discuss an off-lattice model that does not follow the 
G\=o prescription. Using this model, we perform extensive 
folding simulations for a small helical protein. The force field of
the model is simple and based on hydrogen bonds and effective 
hydrophobicity forces (no explicit water). There exist other non 
G\=o-like models with more elaborate force fields that have 
been used for structure prediction with some 
success~\cite{Lee:99,Pillardy:00,Hardin:00}. However, it is 
unclear what the dynamical properties of these models are. 

The original version of our model was presented in Ref.~\cite{Irback:00}
and has three types of amino acids: hydrophobic, polar and glycine. 
This version was applied to a designed three-helix-bundle 
protein with 54 amino acids~\cite{Irback:00}. For a suitable
relative strength of the hydrogen bonds and hydrophobicity forces,
it was found that this sequence does form a stable three-helix 
bundle, except for a twofold topological degeneracy, and that its 
folding transition is first-order-like and coincides with the collapse 
transition (the parameter $\sigma$ of Ref.~\cite{Klimov:98} is zero). 

Here, we extend this model from three to five amino acid 
types, by taking alanine to be intermediate in 
hydrophobicity between the previous two hydrophobic and
polar classes, and by introducing a special geometric 
representation for proline, which is needed to be able to mimic 
the helix-breaking property of this amino acid. 
Otherwise, the model is the same as before. 
The modified model is tested on a real three-helix-bundle
protein, the 10--55-amino acid fragment 
of the B domain of staphylococcal protein A.
The structure of this protein has been 
determined by NMR~\cite{Gouda:92}, and an energy-based structure 
prediction method has been tested on the sequence~\cite{Lee:99}. 
The folding properties have been studied too, both 
experimentally~\cite{Bottomley:94,Bai:97} and theoretically~\cite{Zhou:97,
Zhou:99,Shea:99,Boczko:95,Guo:97,Kolinski:98}. In particular, this  
means that we can compare the behavior of previous G\=o-like models 
to that of our more realistic model.  

\section{Materials and Methods}\label{sec:mod}

\subsection{Geometry}
 
\begin{figure}
\vspace{0mm}
\begin{center}
\epsfig{figure=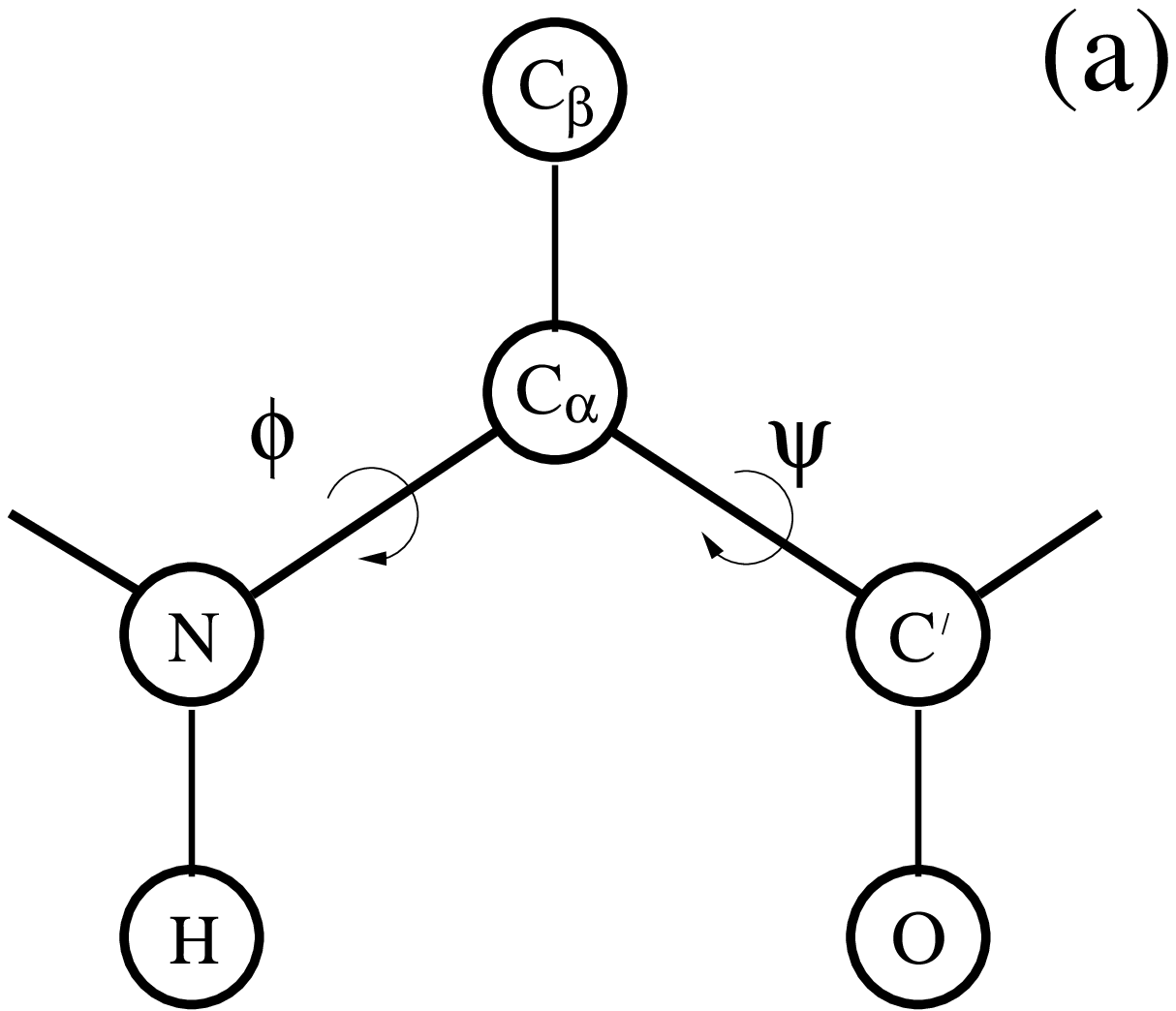,width=6cm,height=5.4cm}
\hspace{10mm}
\epsfig{figure=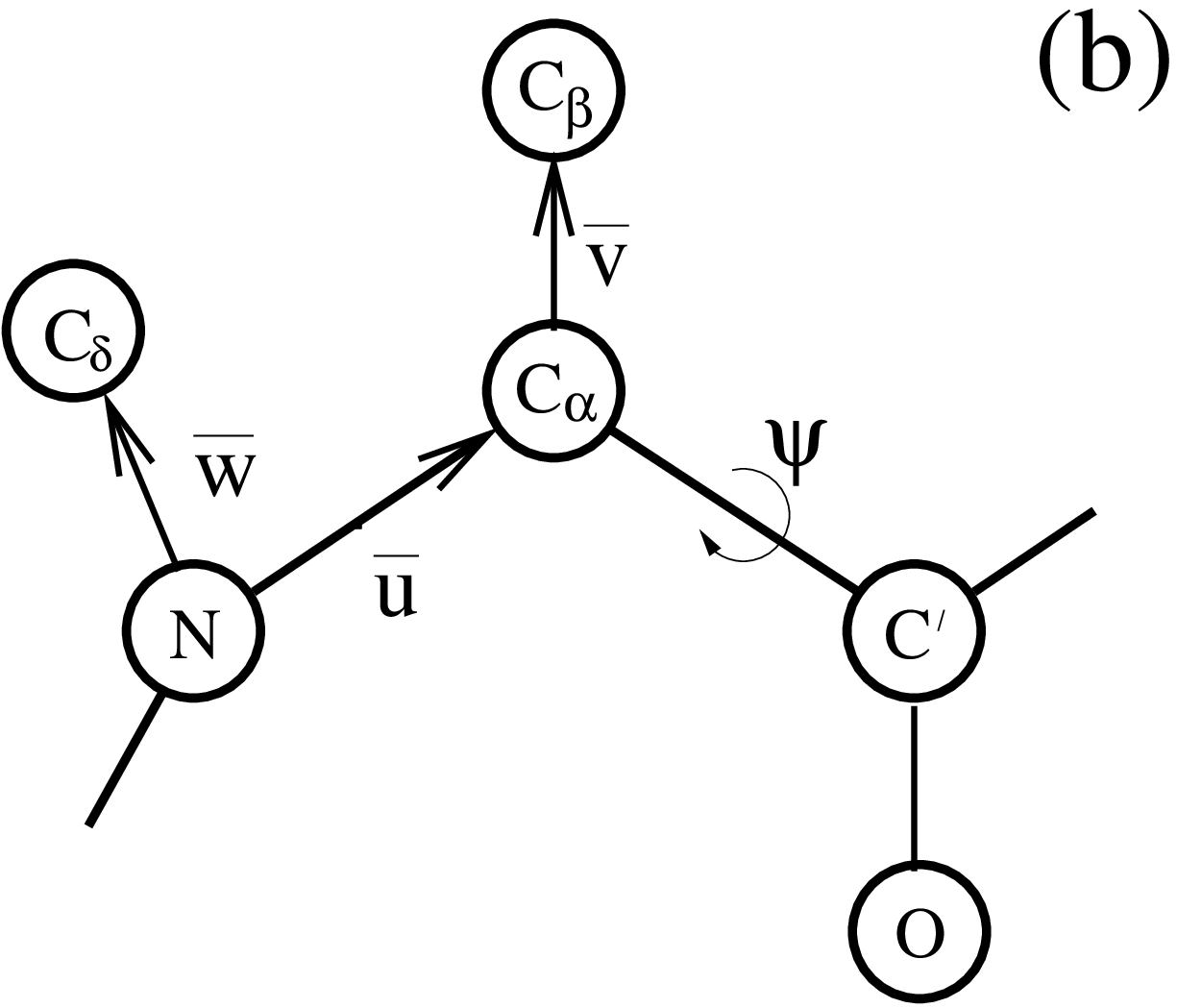,width=6cm,height=5.4cm}
\end{center}
\caption{(a) Schematic figure showing the common geometric
representation for all amino 
acids except glycine and proline. 
(b) The representation of proline. The \Cd\ atom is 
assumed to lie in the plane of the N, \Ca\ and \Cb\ atoms. The N-\Cd\ bond 
vector $\bar w$ is given by $\bar w=-0.596\bar u+0.910\bar v$, where the
vectors $\bar u$ and $\bar v$ are defined in the figure. The 
numerical factors were obtained by an analysis of structures
from the Protein Data Bank (PDB)~\cite{Bernstein:77}.} 
\label{fig:1}
\end{figure}

Our model is an extension of that introduced in Ref.~\cite{Irback:00}. 
It uses three different amino acid representations: one for glycine, 
one for proline and one for the rest. The non-glycine, non-proline 
representation is illustrated in Fig.~\ref{fig:1}a, and is identical to 
that of hydrophobic and polar amino acids in the original model. 
The three backbone atoms N, \Ca\ and \Cp\ 
are all included, whereas the side chain is represented 
by a single atom, a large \Cb. The remaining two 
atoms, H and O, are used to define hydrogen bonds. 
The representation of glycine is the same  
except that \Cb\ is missing. 

The representation of proline is new compared to the original model. 
The side chain of proline is attached to the backbone not only at \Ca, 
but also at N. A well-known consequence of this is that  
proline can act as a helix breaker. For the model to 
be able to capture this important property, we introduce  
a special representation for proline, which is  
illustrated in Fig.~\ref{fig:1}b. It differs from that in
Fig.~\ref{fig:1}a  in two ways: first, the Ramachandran angle 
$\phi$ is held constant, at $-65^\circ$; and second, the H atom 
is replaced by a side-chain atom, \Cd. This more realistic 
representation of proline is needed when studying 
the protein A fragment which has one proline at each of
the two turns.  

All amino acids except proline have the Ramachandran 
torsion angles $\phi$ and $\psi$ (see Fig.~\ref{fig:1}a)   
as their degrees of freedom, whereas $\psi$ is the only   
degree of freedom for proline. All bond lengths, bond angles and
peptide torsion angles ($180^\circ$) are held fixed. Numerical values of 
the bond lengths and bond angles can be found in Ref.~\cite{Irback:00} and 
Fig.~\ref{fig:1}b.

\begin{figure}
\vspace{0mm}
\begin{center}
\epsfig{figure=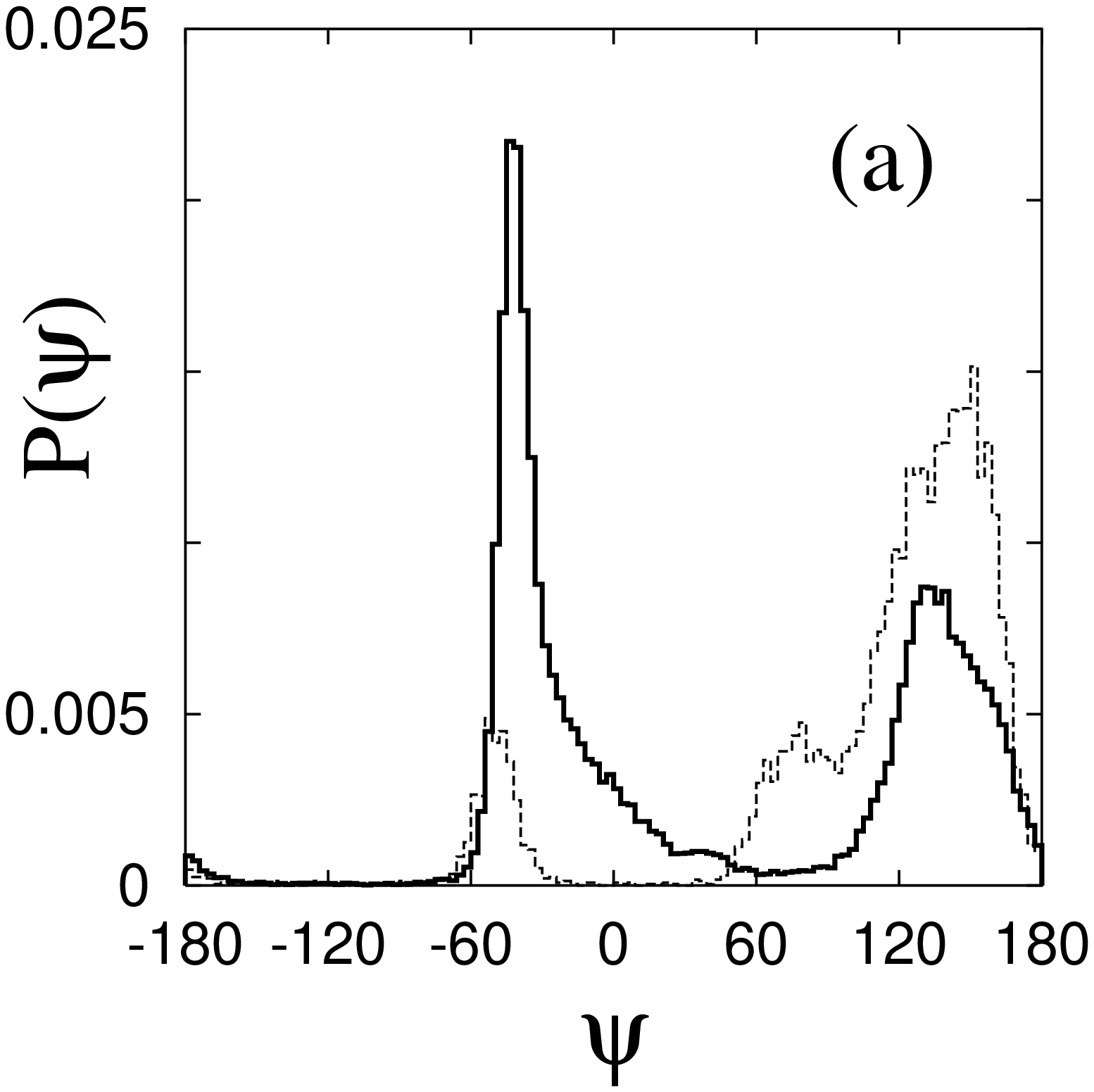,width=6cm}
\hspace{10mm}
\epsfig{figure=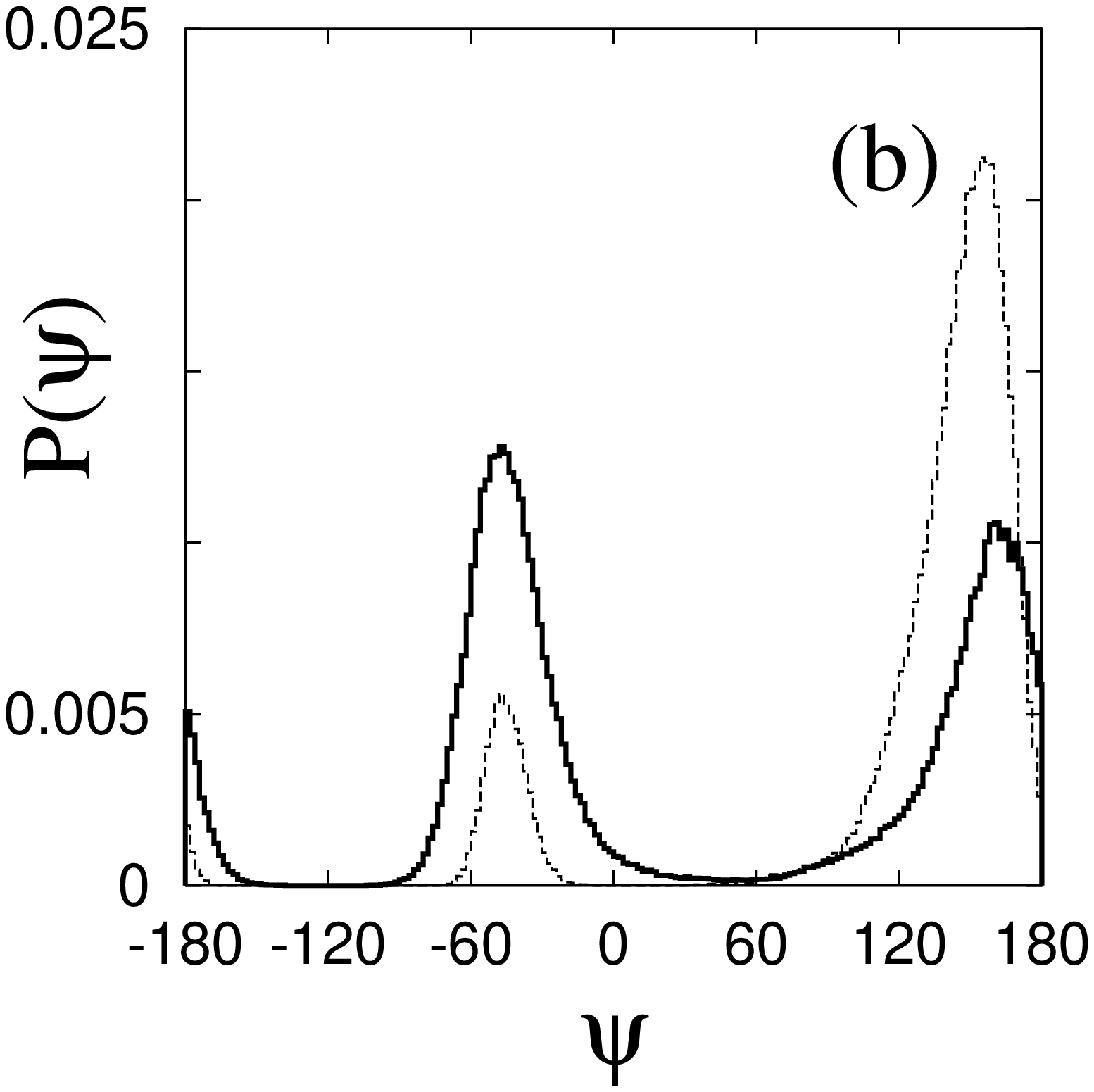,width=6cm}
\end{center}
\caption{(a) Distributions of the Ramachandran angle $\psi$,
based on PDB data. The full (dashed) line represents 
non-glycine, non-proline amino acids that are followed 
by a non-proline (proline) in the sequence. 
(b) The corresponding histograms for the model, as obtained by 
simulations of Gly-X-X (full line) and Gly-X-Pro (dashed line) at $kT=0.55$, 
where X denotes polar amino acids (shown is the $\psi$ distribution  
for the middle of the three amino acids).} 
\label{fig:2}
\end{figure}

The helix-breaking property of proline manifests itself clearly 
in the shape of the $\psi$ distribution for amino acids that are
followed by a proline in the sequence (with the proline on their 
\Cp\ side). Helical values of $\psi$ are suppressed for such amino 
acids. This is illustrated in Fig.~\ref{fig:2}a, where the peak on 
the left corresponds to $\alpha$-helix. From Fig.~\ref{fig:2}b, it can be
seen that the model shows a qualitatively similar behavior. 

\subsection{Force Field}
 
Our energy function 
\beq
E=\Eloc+\Esa+\Ehb+\Ecol
\label{e}\eeq
is composed of four terms. The first two terms $\Eloc$ and
$\Esa$ are local $\phi,\psi$ and self-avoidance potentials,
respectively (see Ref.~\cite{Irback:00}). 
The third term is the hydrogen-bond energy $\Ehb$,
which is given by    
\begin{eqnarray} 
\Ehb&=&\ehb \sum_{ij}
\left[5\left(\frac{\shb}{r_{ij}}\right)^{12}-
6\left(\frac{\shb}{r_{ij}}\right)^{10}\right] 
v(\alpha_{ij},\beta_{ij})\label{hb1}\\
v(\alpha_{ij},\beta_{ij})&=&\left\{ 
        \begin{array}{ll}
 \cos^2\alpha_{ij}\cos^2\beta_{ij} & \ \alpha_{ij},\beta_{ij}>90^{\circ}\\
 0                      & \ \mbox{otherwise}\label{hb2}
         \end{array} \right. 
\end{eqnarray}
where $i$ and $j$ represent H and O atoms, respectively, 
and where $r_{ij}$ denotes the HO distance, $\alpha_{ij}$ the NHO 
angle, and $\beta_{ij}$ the HO\Cp\ angle. 

The last term in Eq.~(\ref{e}), the hydrophobicity or 
collapse energy $\Ecol$, has the form  
\beq
\Ecol=\ecol\sum_{i<j}
\Delta(s_i,s_j)\left[
\left(\frac{\scol}{r_{ij}}\right)^{12}-
2\left(\frac{\scol}{r_{ij}}\right)^{6}
\right]\,,
\label{ecol}\eeq
where the sum runs over all possible \Cb\Cb\ pairs and $s_i$ denotes  
amino acid type. To define $\Delta(s_i,s_j)$, we divide the 
amino acids into three classes: hydrophobic (H; Leu, Ile, Phe), 
alanine (A; Ala) and polar (P; Arg, Asn, Asp, Gln, Glu, His, Lys, Pro, Ser, 
Tyr).\footnote{
Cys, Met, Thr, Trp and Val do not occur in the sequence studied.} 
There are then six kinds of \Cb\Cb\ pairs, and
the corresponding $\Delta(s_i,s_j)$ values are taken to be
\beq
\Delta(s_i,s_j)=\left\{ \begin{array}{rl}
1 & \textnormal{for HH and HA pairs}\\
0 & \textnormal{for HP, AA, AP and PP pairs} 
\end{array}\right.
\label{delta}\eeq

The main change in the force field compared to Ref.~\cite{Irback:00}
is that alanine forms its own hydrophobicity class, besides the previous two 
hydrophobic and polar classes. Alanine is taken as intermediate in 
hydrophobicity, meaning that there is a hydrophobic interaction 
between HA pairs but not between AA pairs. In addition, the interaction 
strength $\ecol$ is increased slightly, from 2.2 to 
2.3.\footnote{The energy unit is dimensionless and such that $k\Tc=0.62$, 
$\Tc$ being the collapse temperature (see Sec.~\protect\ref{sec:res}).}       
Finally, in the self-avoidance potential, 
the \Cd\ atom of proline is assigned the same size as \Cb\ atoms. 
Otherwise, the entire force field, including parameter values, is
exactly the same as in Ref.~\cite{Irback:00}. 

With these changes in geometry and force field, we end up with  
five different amino acid types in the new model. First, we have 
hydrophobic, alanine and polar which share the same geometric
representation but differ in hydrophobicity, and then  
glycine and proline with their special geometries. 

In this paper, we test this model on the 10--55-amino acid fragment 
of the B domain of staphylococcal protein A. Calculated structures are  
compared to the minimized average NMR structure~\cite{Gouda:92}
with PDB code 1bdd. Throughout the paper, this structure is
referred to as the native structure.   
  
As a first test of our model, two different fits to the native 
structure were made. The first fit is purely geometrical.
Here, we simply minimized the root-mean-square deviation (rmsd) 
from the native structure, $\delta$ (calculated over all backbone atoms). 
This was done by using simulated annealing, 
and the best result was $\delta=0.14$~\AA.
In the second fit, we took into account 
the limitations imposed by the first three terms of 
the potential, by minimizing the function 
\beq
\tilde E=\Eloc+\Esa+\Ehb+
\kappa \sum_{i} ({\mathbf r}_{i}-{\mathbf r}_{i}^{0})^2\,,
\label{eaux}\eeq
where $\kappa=1$~\AA${}^{-2}$ and $\{{\bf r}^0_i\}$ denotes 
the structure obtained from the first fit. The minimum-$\tilde E$     
structure had $\delta=0.32$~\AA. These results show that our 
model, in spite of relatively few degrees of freedom,
permits a quite accurate description of the real structure. 

\subsection{Numerical Methods} 

To simulate the thermodynamic behavior of this model, we use simulated 
tempering~\cite{Lyubartsev:92,Marinari:92,Irback:95}, which means that 
the temperature is a dynamical variable (for details, see 
Refs.~\cite{Lyubartsev:92,Marinari:92,Irback:95}). The temperature update is
a standard Metropolis step. Our conformation updates are of two 
different types: the simple non-local pivot move where a single 
torsion angle is turned, and the semi-local biased Gaussian step 
proposed in Ref.~\cite{Favrin:01}. The latter method works with the 
Ramachandran angles of four adjacent amino acids. These are turned 
with a bias toward local rearrangements of the chain. The degree of 
bias is governed by a parameter $b$. In our thermodynamic simulations, 
we take $b=10$ (rad/\AA)${}^2$, which gives a strong bias toward 
deformations that are approximately local~\cite{Favrin:01}.

\begin{figure}
\vspace{0mm}
\begin{center}
\epsfig{figure=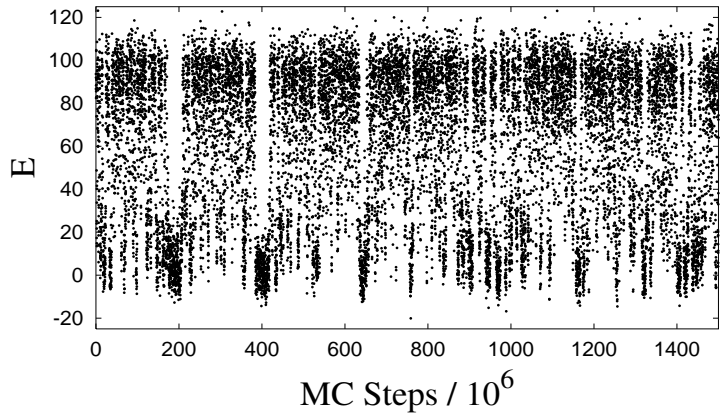,width=10cm,height=5.7cm,angle=0}
\caption{Monte Carlo evolution of the energy in a simulated-tempering run.}  
\label{fig:3}
\end{center}
\end{figure}

Figure~\ref{fig:3} shows the evolution of the energy in a 
simulated-tempering run that took about two weeks on an 800~MHz processor. 
Data corresponding to all the different temperatures are 
shown (eight temperatures, ranging from $kT=0.54$ to $kT=0.90$). 
We see that there are many independent visits to low-energy states, which is
necessary in order to get a reliable estimate of the relative populations of 
the folded and unfolded states. To test the usefulness of the semi-local
update, we repeated the same calculation using pivot moves only. 
The difference in performance was not quantified, but it was clear       
that the sampling of low energies was less efficient in the run 
relying solely on pivot moves.  

For our kinetic simulations, we do not use the pivot update but
only the semi-local method. The parameter $b$ is taken to be
1~(rad/\AA)${}^2$ in the kinetic runs, which turned out to give an 
average change  in the end-to-end vector squared of about 0.5~\AA${}^2$. 

\newpage

\section{Results and Discussion}\label{sec:res}

\subsection{Thermodynamics}\label{sec:thermo}

\begin{figure}
\vspace{0mm}
\begin{center}
\epsfig{figure=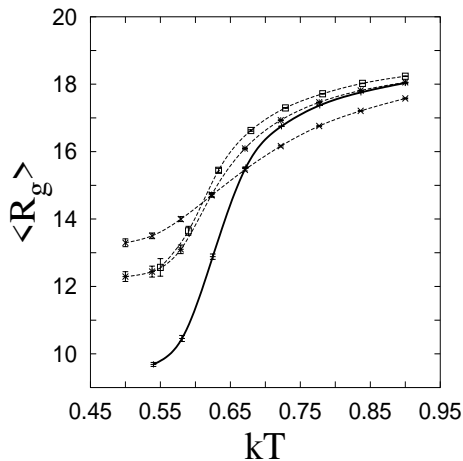,width=6cm,height=6cm,angle=270}
\end{center}
\caption{The radius of gyration (in \AA) against temperature.
Full and dashed lines represent the protein A sequence and 
the three random sequences (see the text), respectively.}
\label{fig:4}
\end{figure}

We begin our study of the model defined in Sec.~\ref{sec:mod}
by locating the collapse transition. In Fig.~\ref{fig:4}, we show 
the radius of gyration (calculated over all backbone atoms)
against temperature for both the protein A sequence and three 
random sequences with the same length and composition.
The random sequences were generated keeping the two prolines of the 
protein A sequence fixed at their positions, one at each turn. 
The remaining 44 amino acids were randomly reshuffled.

Naively, one may expect these sequences to show similar collapse
behaviors, since the composition is the same. However, the protein A 
sequence turns out to collapse much more efficiently than the 
random sequences (see Fig.~\ref{fig:4}). The native structure has 
a radius of gyration of 9.25~\AA, which is significantly smaller than 
one finds for the random sequences in this temperature range. 
The specific heat (data not shown) has a pronounced peak
in the region where the collapse occurs. Taking the maximum 
as the collapse temperature $\Tc$, we obtain $k\Tc=0.62$ for the 
protein A sequence. 

The chain collapse is not as abrupt for the protein A sequence
as for the designed sequence studied in Ref.~\cite{Irback:00}. 
This is not surprising, as that sequence has a hydrophobicity
pattern that fits its native structure perfectly. The protein
A sequence does not have a fully perfect hydrophobicity pattern,      
but still the collapse behavior is highly cooperative, as can be seen 
from the comparison with the random sequences.  

Next, we turn to the structure of the collapsed state. 
As a measure of similarity with the native structure, 
we use  
\beq
Q=\exp(-\delta^2/100~{\rm \AA}^2)\,,
\label{Q}\eeq
where $\delta$, as before, denotes rmsd. An alternative  
would be to base the similarity measure on the number of 
native contacts present, rather than rmsd. The problem with 
such a definition is that it does not provide an efficient 
discrimination between the two possible topologies of a three-helix 
bundle~\cite{Wallin:01} --- the third helix can be either in front of
or behind the U formed by the first two helices.
This problem is avoided by using rmsd.

\begin{figure}[t]
\vspace{0mm}
\begin{center}
\epsfig{figure=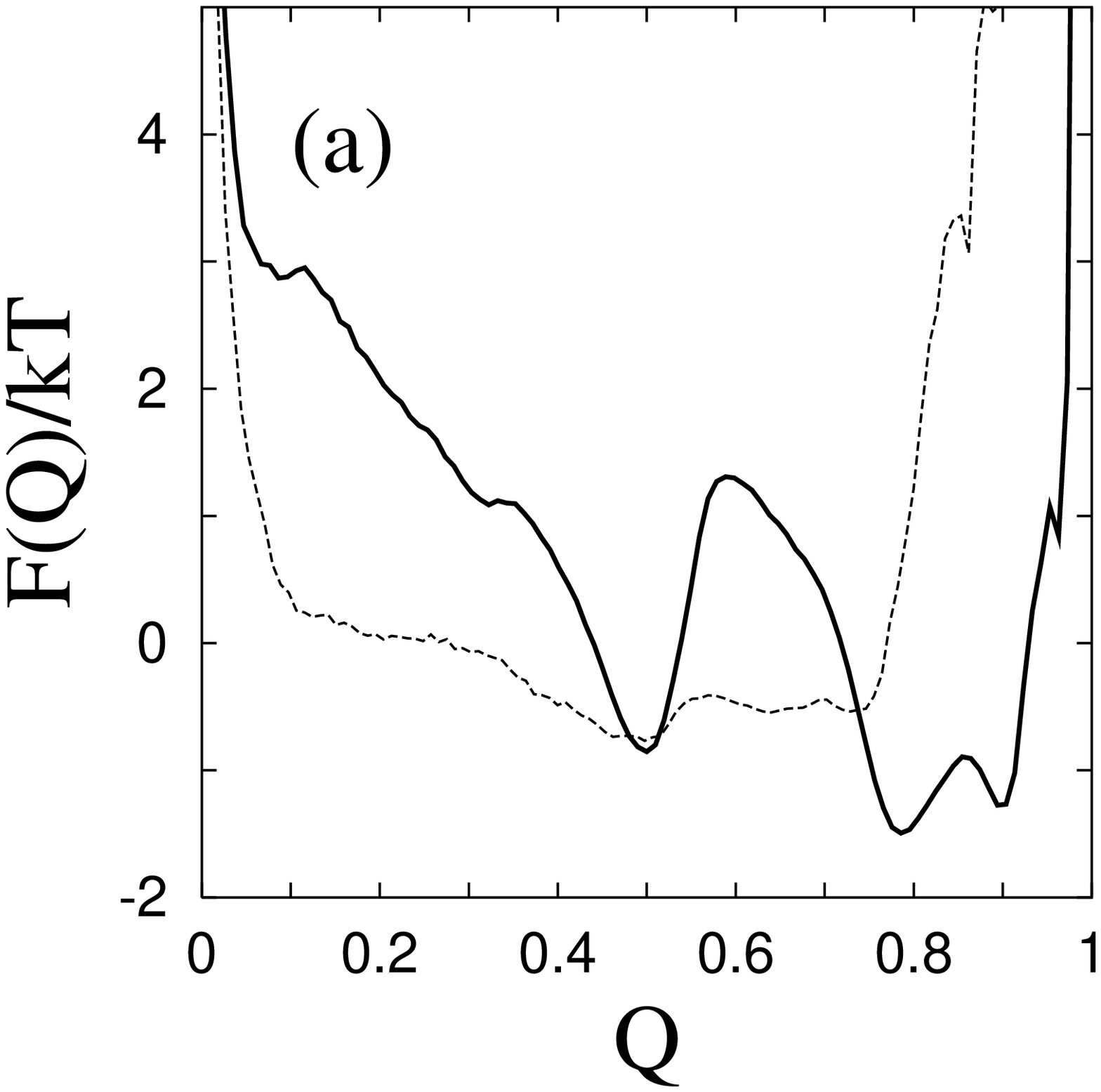,width=6cm,angle=270}
\hspace{10mm}
\epsfig{figure=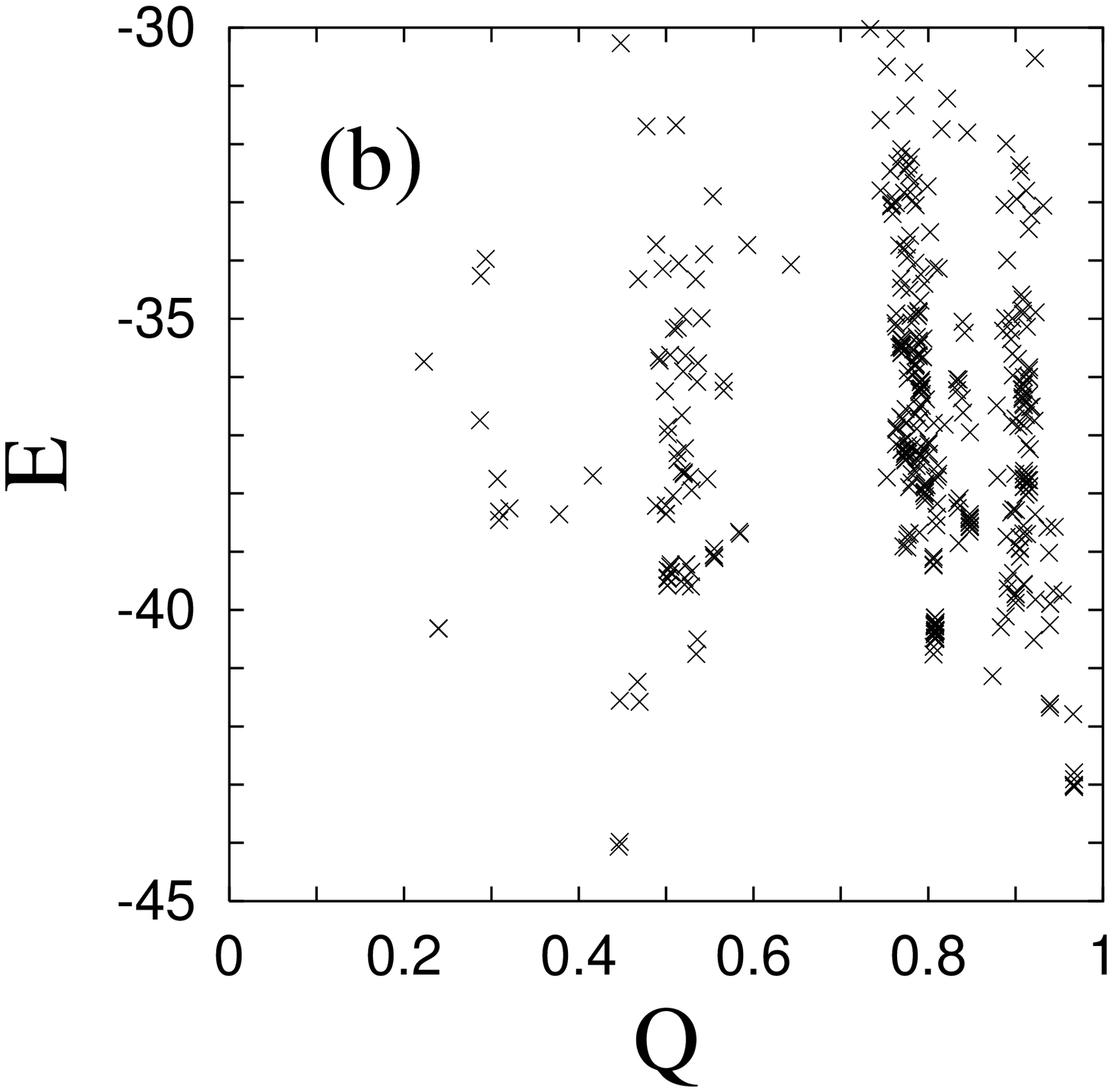,width=6cm,angle=270}
\end{center}
\caption{(a) Free-energy profile $F(Q)=-kT\ln P(Q)$ at $kT=0.54$ (full line),
where $P(Q)$ is the probability distribution of $Q$. Also shown 
(dashed line) is the result for one of the random sequences at $kT=0.50$.
(b) $Q,E$ scatter plot for quenched conformations with low energy.}
\label{fig:5}
\end{figure}

In Fig.~\ref{fig:5}a, we show the free-energy profile $F(Q)$ in the
collapsed phase at $kT=0.54$. 
We see that there is a broad minimum 
at $Q\approx0.8$--0.9, with two distinct local minima at   
$Q=0.78$ and $Q=0.90$, respectively. Both these
minima correspond to the native overall topology. There is also a 
minimum at $Q=0.50$, which corresponds to
the wrong topology. The $Q=0.50$ minimum is more
narrow and slightly higher, so the native topology is the
favored one. However, it should be stressed that it is difficult
to discriminate between the two topologies using a pairwise 
additive potential (see Sec.~\ref{sec:f-t}). To be able to do that
in a proper way, it is likely that one has to include multibody
terms and/or more side-chain atoms in the model. 

The main difference between the two minima at $Q=0.78$ and $Q=0.90$ 
lies in the shape and orientation of helix III, 
which comprises amino acids 41--55 in the native structure. At the 
$Q=0.78$ minimum, there tends to be a sharp bend in this segment, 
and the amino acids before the bend, 41--44, are disordered rather 
than helical. The remaining amino acids, 45--55, tend to make a helix, 
but its orientation differs from that in the native structure. 
Relative to the $Q=0.90$ minimum, where helix III is much more 
native-like, we find that the $Q=0.78$ minimum is entropically favored 
but energetically disfavored. The separation in energy between  
these minima is probably underestimated by our model. There is, 
for example, a stabilizing electrostatic interaction between helices 
I and III in the native structure (Glu16-Lys50), which should favor
the $Q=0.90$ minimum but is missing in our model. 

Also shown in Fig.~\ref{fig:5}a is the result for one of the 
random sequences. The probability of finding this sequence in
the vicinity of the native structure is, not unexpectedly, very low.  
The same holds true for the other two random sequences 
too (data not shown).  

\begin{figure}
\vspace{0mm}
\begin{center}
\epsfig{figure=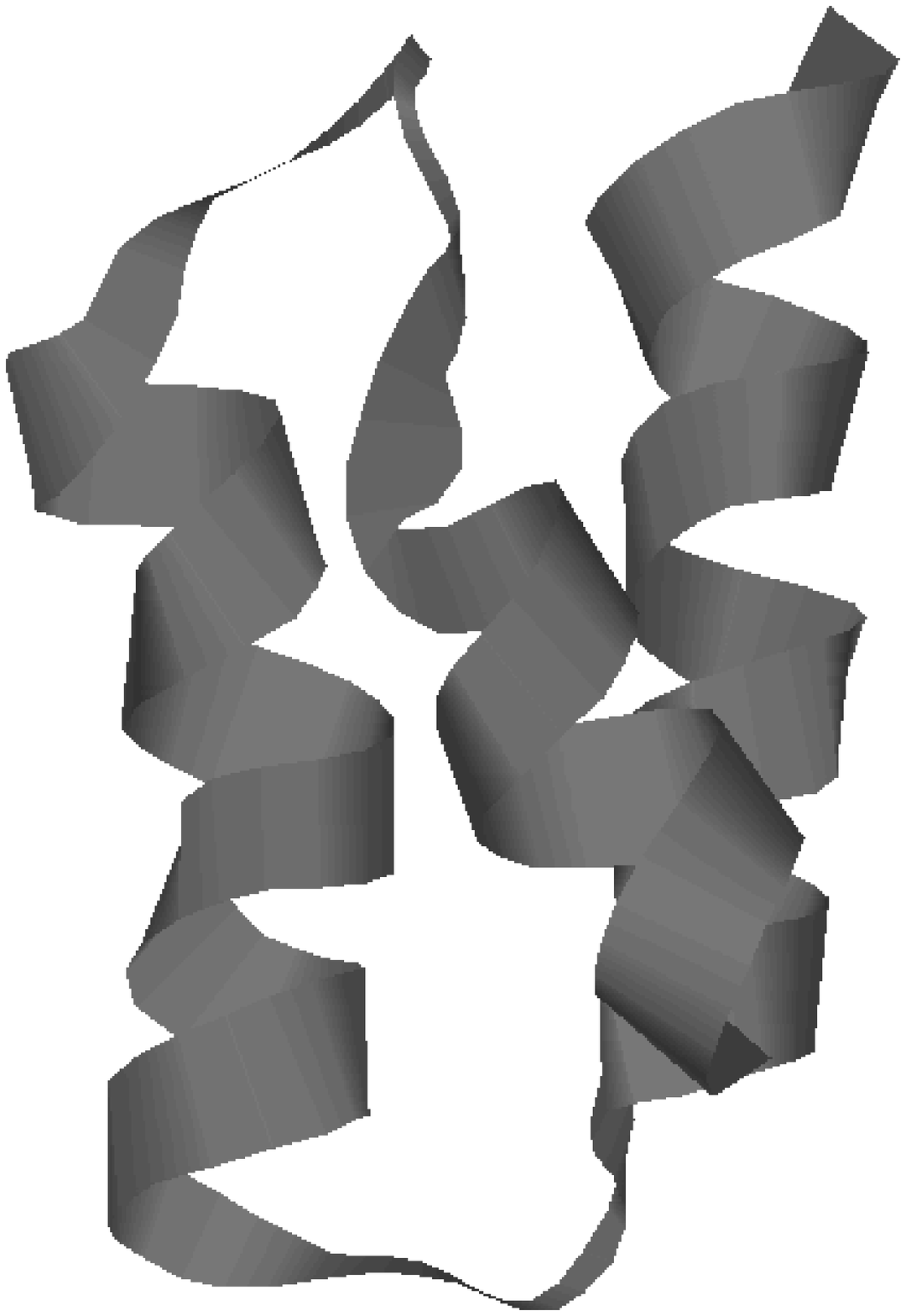,width=5cm}
\hspace{25mm}
\epsfig{figure=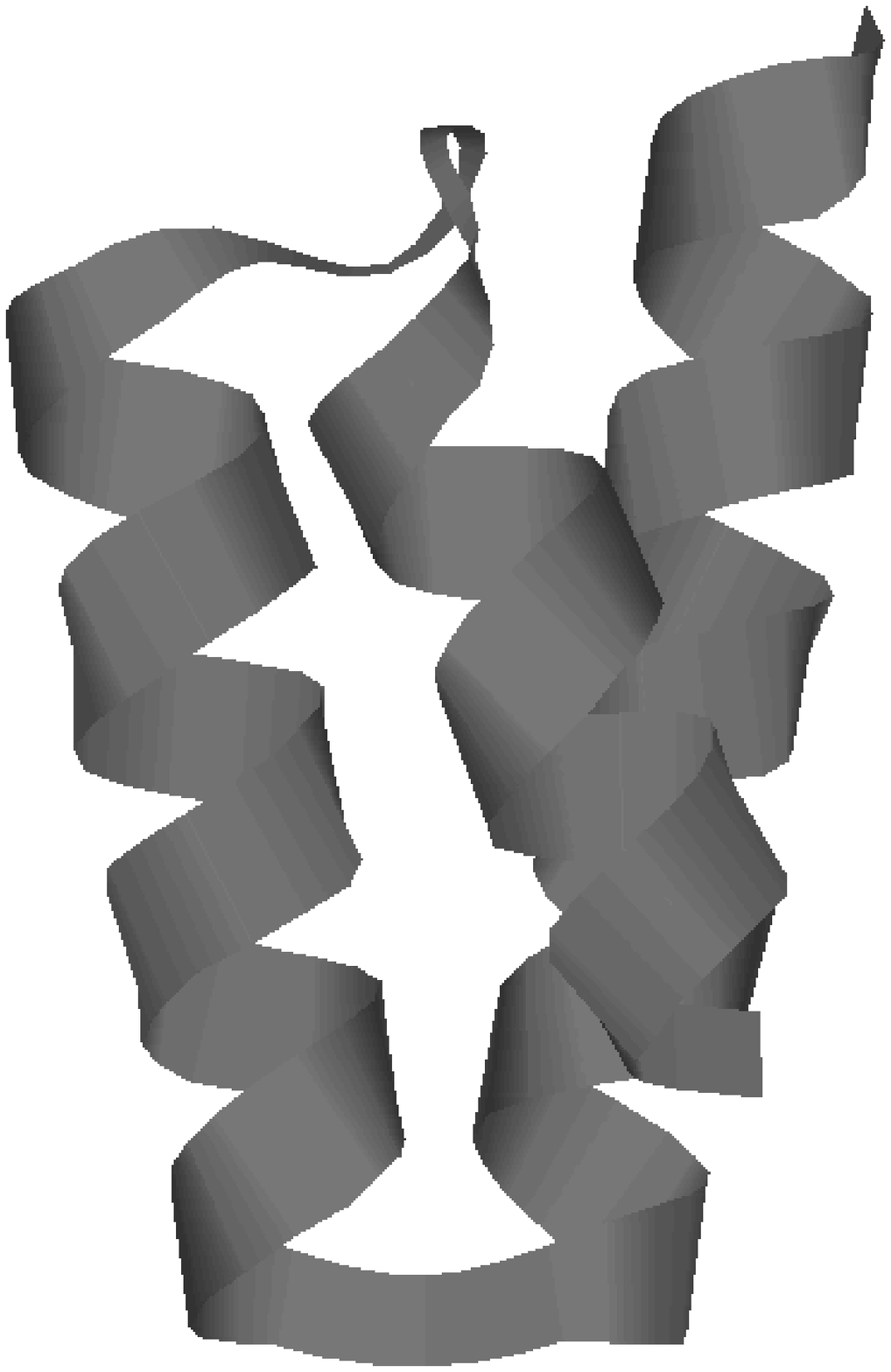,width=5cm}
\end{center}
\caption{Schematic illustrations of the native structure (left) and our 
minimum-energy structure for the native topology (right). 
Drawn with RasMol~\cite{Sayle:95}.}
\label{fig:6}
\end{figure}

To extract representative conformations for the collapsed
state, we used simulated annealing followed by a  
conjugate-gradient minimization. Using this procedure, a large  
set of low-temperature Monte Carlo conformations were quenched to
zero temperature. In Fig.~\ref{fig:5}b, we show the quenched 
conformations with lowest energy in a $Q,E$ scatter plot. Our
minimum-energy structure is found at $Q=0.44$, corresponding to  
$\delta=9.1$~\AA. However, our thermodynamic calculations 
show that this conformation is not very relevant, in spite of 
its low energy. If we restrict ourselves to conformations with the 
native-like and thermodynamically most relevant topology, then 
the lowest energy is at $Q=0.97$, corresponding to $\delta=1.8$~\AA.   
This conformation is shown in Fig.~\ref{fig:6} along with the
native structure. It is worth noting that the
$Q=0.44$ and $Q=0.97$ minima both were revisited in independent
runs.  

These results can be compared with those of Scheraga and 
coworkers~\cite{Lee:99}, 
who tested an energy-based structure prediction method on the same 
sequence. With their energy function, the global minimum was found to have 
an rmsd of 3.8~\AA\ from the native structure (calculated over \Ca\ atoms).

\subsection{Helix Stability}

Having discussed the overall thermodynamic behavior, we 
now take a closer look at the stability of the secondary structure
and how it varies along the chain. To this end, we  
monitored the hydrogen-bond energy  
between the CO group of amino acid $i$ and the NH group of amino acid $i+4$
[see Eqs.~(\ref{hb1},\ref{hb2})], $\Eihb(i)$, as a function of $i$.
This was done not only for the protein A sequence, but also 
for the corresponding three one-helix segments, which are 
listed in Table~\ref{tab:1}. An experimental study~\cite{Bai:97} of 
essentially the same three segments found segment III to be the
only one that shows some stability on its own.  

\begin{table}[t]
\begin{center}
\begin{tabular}{clc}
Segment & Sequence               & Amino acids \\ \hline
I       & QQNAFYEILHL            & 10--20 \\
II      & NEEQRNGFIQSLKDD        & 24--38 \\
III     & QSANLLAEAKKLNDA        & 41--55 
\end{tabular}
\caption{The one-helix fragments studied.}
\label{tab:1}
\end{center}
\end{table}

\begin{figure}[t]
\vspace{0mm}
\begin{center}
\epsfig{figure=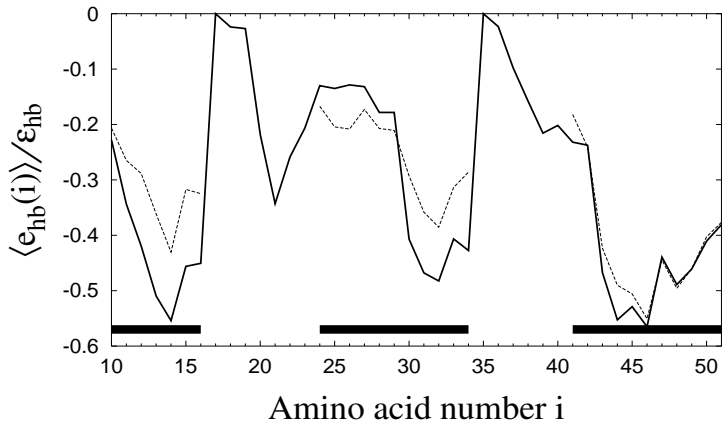,width=10cm,height=5.7cm}
\end{center}
\caption{Hydrogen-bond profile showing the normalized average  
energy of $\alpha$-helical hydrogen bonds, $\ev{\Eihb(i)}/\ehb$, against  
amino acid number $i$, at $kT=0.58$. The full line represents the
protein A sequence, whereas the dashed lines represent the
corresponding three one-helix segments (see Table~\ref{tab:1}). The thick 
horizontal lines indicate hydrogen bonds present in the 
native structure.}
\label{fig:7}
\end{figure}

The results of our calculations are shown in Fig.~\ref{fig:7},
from which we see that the difference between the full sequence
and the one-helix segments is not large in the model. However,
the segments I and II definitely make less stable helices
on their own than as interacting parts of the full system;
they are stabilized by interhelical interactions. Furthermore,
among the three one-helix segments, the model correctly predicts
segment III to be the most stable one. That this segment does 
not get more stable as part of the full system is probably 
related to the observation above that helix III is distorted
at the $Q=0.78$ minimum.  

A striking detail in Fig.~\ref{fig:7} is that the beginning 
of segment II is quite unstable. This can be easily understood.   
This segment has a flexible glycine at position 30, and the amino acids 
before the glycine, 24--29, are all polar, so there are   
no hydrophobic interactions that can help to stabilize this part.  

\subsection{Kinetics}

\begin{figure}
\vspace{0mm}
\begin{center}
\epsfig{figure=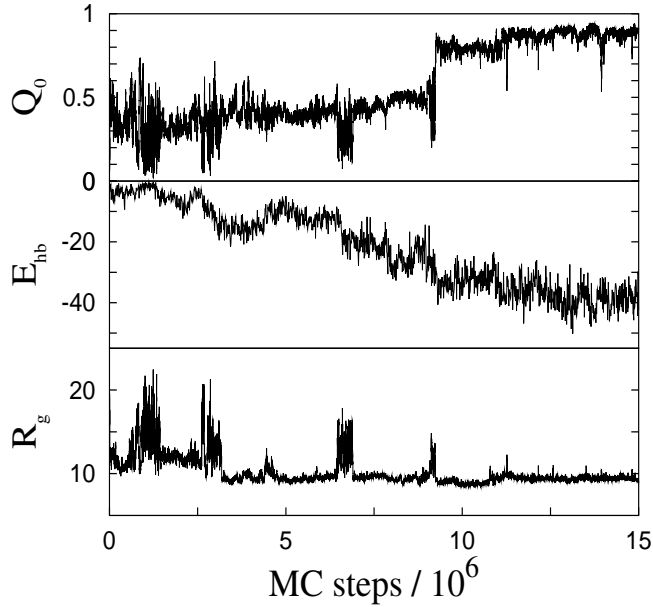,width=8cm,height=8.5cm,angle=270}
\end{center}
\caption{Monte Carlo evolution of the similarity parameter $Q_0$ (top), 
the hydrogen-bond energy $\Ehb$ (middle) and the radius of gyration 
$\Rg$ (bottom) in a kinetic simulation at $kT=0.54$.}
\label{fig:8}
\end{figure}

Using the semi-local update~\cite{Favrin:01}, we performed
a set of 30 kinetic simulations at $kT=0.54$. The runs were 
started from random coils. There are big differences between
these runs, partly because the system, as it should, sometimes
spent a significant amount of time in the wrong topology.
Nevertheless, the data show one stable and interesting trend,
namely, that the formation of helices was never 
faster than the collapse. This is illustrated in 
Fig.~\ref{fig:8}, which shows the evolution of the similarity
parameter $Q_0$, the hydrogen-bond energy $\Ehb$ and the 
radius of gyration, $\Rg$, in one of the runs. $Q_0$ is  
defined as $Q$ in Eq.~(\ref{Q}), except that it measures 
similarity to the optimized model structure in Fig.~\ref{fig:6} 
rather than the native structure. In Fig.~\ref{fig:8}, we see that $\Ehb$ 
converges slowly, whereas the collapse occurs relatively early.

Now, at a first glance, it may seem easy to make the helix formation 
faster by simply increasing the strength of the hydrogen bonds. Therefore,
it is important to note that the hydrogen bonds cannot be made
much stronger without making the ground state non-compact and
thus destroying the three-helix bundle~\cite{Irback:00b}. This means
that the conclusion that the collapse is at least as fast as 
helix formation holds for any reasonable choice of parameters in this 
model.  

It is interesting to compare these results to those of
Zhou and Karplus~\cite{Zhou:99}, who studied the same protein
using a  G\=o-type potential and observed fast folding when
the G\=o forces were strong. Under these conditions, the helix
formation was found to be fast, whereas the collapse was the
rate-limiting step. 

However, a G\=o-like model ignores a large fraction of the 
interactions that drive the collapse, which can make the 
collapse artificially slow. In a recent G\=o model study~\cite{Shimada:01}, 
this problem was addressed by eliminating backbone terms from 
the potential until a reasonable helix stability was achieved. 
No such calibration was carried out in Ref.~\cite{Zhou:99}.  
This may explain why these authors find a behavior that 
our model cannot reproduce. 

Let us finally mention that we also performed the same type of
kinetic simulations for the designed sequence studied in 
Ref.~\cite{Irback:00} which, as discussed earlier, has a very 
abrupt collapse transition. It turns out that $\Ehb$ and $\Rg$ 
evolve in a strongly correlated manner in this case. So, the helix 
formation and collapse occur simultaneously for this sequence.
  
\subsection{Fine-tuning?}\label{sec:f-t}

In Sec.~\ref{sec:thermo}, we discussed the relative weights 
of the two possible overall topologies, which is a delicate
issue. What changes are needed in order for the model to
more strongly suppress the wrong topology? Is it necessary
to change the form of the energy function, or would it be
sufficient to fine-tune the interaction matrix  
$\Delta(s_i,s_j)$ in Eq.~(\ref{ecol})?

One way to do such a fine-tuning of $\Delta(s_i,s_j)$ 
would be to maximize $\ev{Q}'$, where $Q$ is the similarity
parameter and $\ev{\cdot}'$ denotes a thermodynamic average 
restricted to compact conformations ($\Rg<10$~\AA\ say). 
This is essentially the overlap method of Ref.~\cite{Bastolla:00}.
The gradient of the quantity $\ev{Q}'$ can be written as 
\beq
\frac{\partial\ev{Q}'}{\partial\Delta(s_i,s_j)}=
-\frac{\ecol}{kT}\left(\ev{QX}'-\ev{Q}'\ev{X}'\right)\,,
\label{grad}\eeq
where $X$ is a sum of Lennard-Jones terms, 
$(\scol/r_{ij})^{12}-2(\scol/r_{ij})^6$, over all possible
\Cb\Cb\ pairs of type $s_i,s_j$. 

We calculated the $Q,X$ correlation in Eq.~(\ref{grad}) 
for all pairs $s_i,s_j$ with
$\Delta(s_i,s_j)=1$ at $kT=0.54$, and found that 
$|\partial\ev{Q}'/\partial\Delta(s_i,s_j)|$ 
was small ($\le0.15$) for all these pairs. Hence, there is no 
sign that a significant increase in $\ev{Q}'$ can be achieved by  
fine-tuning $\Delta(s_i,s_j)$; the contact patterns seem to be 
too similar in the two topologies. To include
more side-chain atoms and/or multibody terms in the model 
is likely to be a more fruitful approach.  
  
\section{Conclusion}

We have explored a five-letter protein model with five to 
six atoms per amino acid, where the formation of native 
structure is driven by hydrogen bonding and effective hydrophobicity 
forces. This model, which does not follow the G\=o prescription,  
was tested on a small but real sequence, a three-helix-bundle 
fragment from protein A.

Using this model, the protein A sequence was found 
to collapse much more efficiently than random sequences with 
the same composition. In the collapsed phase, we found that 
the native topology dominates, although the suppression 
of the wrong three-helix-bundle topology is not strong. 
Energy minimization constrained to the thermodynamically    
favored topology gave a structure with an rmsd of 1.8~\AA\ from
the native structure. 
   
In our kinetic simulations, the collapse was always at least as
fast as helix formation, which is in sharp contrast with previous 
results for the same protein that were obtained using a G\=o-like 
\Ca\ model~\cite{Zhou:99}. A possible explanation for the 
conflicting conclusions is that the G\=o approximation makes the collapse
artificially slow by ignoring a large fraction of the interactions
driving the collapse. In our model, the conclusion that the helix formation
is not faster than collapse seems unavoidable; if one tries to speed 
up the helix formation by increasing the strength of the hydrogen bonds, 
then the chain does not fold into a compact helical bundle. 

The force field of our model was deliberately kept simple. In particular,
the hydrophobicity potential was taken to be pairwise additive,
with a simple structure for the interaction matrix $\Delta(s_i,s_j)$
[see Eq.~(\ref{delta})]. In the future, it would be very interesting 
to look into the behavior of the model in the presence of multibody terms. 
A simpler alternative is to stick to the pairwise additive potential 
and fine-tune the parameters $\Delta(s_i,s_j)$. However, the calculations 
in this paper give no indication that there is much to be gained from 
such a fine-tuning.

\subsection*{Acknowledgments}

This work was in part supported by the Swedish Foundation for Strategic 
Research.

\end{document}